\def\gsim{~\rlap{$>$}{\lower 1.0ex\hbox{$\sim$}}}
\def\lsim{~\rlap{$<$}{\lower 1.0ex\hbox{$\sim$}}}

\def\h2o{\rm{H_{2}O}}
\def\mh2{\rm{H_{2}}}

\def\co2{\rm{CO_{2}}}
\def\ch4{\rm{CH_{4}}}
\def\n2{\rm{N_{2}}}
\def\teff{\rm{T_{eff}}}
\documentclass[12pt, preprint]{aastex}
\usepackage{graphicx}
\usepackage{natbib,color,lscape}
\usepackage{subfigure}
\usepackage{threeparttable}
\usepackage{url}
\usepackage{threeparttable}
\usepackage{rotating}

\begin{document}
\title{HABITABLE MOIST ATMOSPHERES ON TERRESTRIAL PLANETS NEAR THE INNER EDGE OF THE HABITABLE ZONE AROUND M-DWARFS }
\author{Ravi kumar Kopparapu\altaffilmark{1,2,3,4}, 
        Eric T. Wolf\altaffilmark{5}
        Giada Arney\altaffilmark{1,3},
        Natasha E. Batalha\altaffilmark{1,7, 8}
        Jacob Haqq-Misra\altaffilmark{3,4},
        Simon L. Grimm\altaffilmark{6},
        Kevin Heng\altaffilmark{6}}

\altaffiltext{1}{NASA Goddard Space Flight Center, 8800 Greenbelt Road, Mail Stop 699.0
Building 34, Greenbelt, MD 20771} 
\altaffiltext{2}{Department of Astronomy, University of Maryland College Park, College Park, MD}
\altaffiltext{3}{NASA Astrobiology Institute's Virtual Planetary Laboratory, P.O. Box 351580, Seattle, WA 98195, USA}
\altaffiltext{4}{Blue Marble Space Institute of Science, 1001 4th Ave, Suite 3201, Seattle, Washington 98154, USA}
\altaffiltext{5}{Department of Atmospheric and Oceanic Sciences, Laboratory for Atmospheric and Space Physics, University of Colorado
Boulder, Boulder, Colorado, USA}
\altaffiltext{6}{Center for Space Habitability, University of Bern, Sidlerstrasse 5, Exact Sciences Building, CH-3012, Bern, Switzerland}
\altaffiltext{7}{Department of Astronomy \& Astrophysics, Penn State University, 
Davey Laboratory, University Park, PA 16802, USA}
\altaffiltext{8}{Center for Exoplanets and Habitable Worlds, The Pennsylvania State University, University Park, PA 168a02, USA}

\begin{abstract}
Terrestrial planets in the habitable zones (HZs) of low-mass stars and cool dwarfs have received
significant scrutiny recently 
 Transit spectroscopy of such planets with {\it JWST} represents our best shot at obtaining the spectrum of a habitable planet
within the next decade. As these planets are likely tidal-locked, improved 3D numerical simulations of such planetary atmospheres 
are needed to guide target selection.
Here we use a 3-D climate system model, updated with new water-vapor absorption coefficients derived from the HITRAN 2012
 database, to study ocean covered planets at the inner edge of the HZ around late-M to mid-K stars ($2600$K $ \le T_{eff} \le 4500K$). Our
results indicate that these updated water-vapor coefficients result in significant warming compared to previous studies, so the 
inner HZ around M-dwarfs is not as close as suggested by earlier work. Assuming synchronously rotating Earth-sized and Earth-massed planets with background 1 bar $\n2$ atmospheres, we find that planets at
the inner HZ of stars with $T_{eff}> 3000$K undergo the
classical ``moist-greenhouse'' ($\h2o$ mixing ratio $> 10^{-3}$ in the stratosphere) at significantly
lower surface temperature ($\sim 280$K) in our 3-D model compared with 1-D climate models ($\sim 340$K).
This implies that some planets around low mass stars can simultaneously undergo water-loss and remain habitable.
However,  for star with $T_{eff} \le 3000$K, 
planets at the inner HZ may directly transition to a runaway state, while bypassing the moist greenhouse water-loss entirely.
We analyze transmission spectra of planets in a moist greenhouse regime, and find that there are several prominent $\h2o$ features, including a 
broad feature between 5-8 microns, within {\it JWST} MIRI instrument range. Thus, relying only upon standard Earth-analog spectra 
with 24-hour rotation period around M-dwarfs for habitability studies
will miss the strong $\h2o$ features that one would expect to see on synchronously rotating planets around M-dwarf stars, 
with {\it JWST}.
 
\end{abstract}
\keywords{planets and satellites: atmospheres}

\maketitle

\section{Introduction}
\label{intro}

Discoveries of terrestrial planets in the habitable zone (HZ) by both the {\it Kepler} mission and ground-based surveys
\citep{Kane2016, Anglada-Escude2016, Gillon2017} have shown that potentially habitable planets are common in our galaxy 
\citep{Petigura2013, FM2014, Silburt2015, DC2015}. With the discovery of the nearest HZ planets around Proxima Centauri
and TRAPPIST-1,  we 
also have an opportunity  to characterize the atmospheres of these planets in the not too distant 
future. There is an extensive literature on modeling and speculating on the potential habitability of terrestrial planets in 
the HZ \citep{Abe2011, Selsis2007b, Wordsworth2010, HK2015, Bolmont2017, Ribas2016, Turbet2016, Meadows2016, Barnes2016}, and 
studies such as these provide a valuable contribution in understanding, and eventually interpreting, any 
biosignatures. In particular, low mass stars (late-K and M dwarf stars) provide our best opportunity for detecting and characterizing 
habitable terrestrial planets in the coming decade.  The small size and compact HZs of these stars allows for a greater chance of detection of 
terrestrial sized planets and the higher frequency of transits for planets in the HZs of low-mass stars (due to their shorter orbital
 periods) means that higher signal-to-noise levels can be observed in less time compared to planets orbiting around hotter stars.

Crucial to interpreting these discoveries is the location of the HZ, where  liquid water remains thermodynamically stable on the planet's
 surface indefinitely. Note, that this particular definition of the HZ is geared towards remote detection of life through observations of 
exoplanet atmospheres, and not a statement on the planets' habitability itself \citep{Kasting2014}. It is also useful in determining 
the occurrence of
potentially habitable planets in our galaxy, as was done with {\it Kepler} data. Several groups
 have studied the limits of the main-sequence HZ \citep{Kasting1993, PG2011, Kopp2013, Leconte2013a, Yang2013, Barnes2013, zsom2013, Kopp2014, WT2014, Yang2014a, Way2015, WT2015, Leconte2015, Godolt2015, Kopp2016, HM2016, RK2017, Salameh2017} using both 1-D and 3-D climate models, and corresponding climate 
transitions that planets undergo at these limits. Many of these models assume water-rich ($\sim 1$ Earth ocean) planets,
which is reasonable if one wants to study the surface habitability of a planet. The validity of this assumption can be
debated, particularly for planets in the HZ of M-dwarf stars due to their high pre-main sequence luminosity 
\citep{RK2014, LB2015, TI2015}, or atmospheric erosion due to high XUV fluxes (Airapetian et al. 2017). But this can only be verified with more 
observations, and more detections, of these kinds of planets.  

Both 1-D and 3-D models show that, for a water-rich planet near the inner edge of the HZ, as the surface temperature increases
due to increased stellar radiation, water vapor becomes a significant fraction of both the
troposphere and stratosphere \citep{Kasting2015}. 
$\h2o$ is a strong absorber both in the longwave and the shortwave.  Thus, increasing atmospheric water vapor limits the maximum
outgoing thermal radiation (Goldblatt et al. 2013), and also increases absorption of solar radiation in the near-IR, thereby 
lowering a planets' albedo.  This effect
is accentuated on planets around M-dwarfs because the radiation from M-dwarfs peaks in the near-infrared 
where there are strong water absorption bands.
Our own previous work on HZs (Kopparapu et al., 2013, 2014) using a 1-D radiative-convective,
cloud-free climate model, and updated with new and stronger $\h2o$ absorption coefficients
from HITRAN2008 and HITEMP2010 line-by-line (LBL) databases, showed that the inner edge of the HZ
can be significantly farther from the star than previous estimates \citep{Kasting1993}.

However, a critical component for discerning the climate of terrestrial planets around M-dwarf stars, and thus the HZ, is to consider the 
effects of slow and synchronous rotation caused by tidal locking.  Planets in the HZ of low mass stars are likely to find themselves in 
synchronous rotation due to strong tidal interactions with the host star \citep{DB2009, Barnes2013}.  Planets in synchronous rotation will have permanent day 
and night hemispheres \citep{Joshi1997}. Slow planetary rotation has a profound impact on the climate of Earth-like planets 
(Merlis \& Schneider 2010; Yang et al. 2013; Yang et al. 2014; Carone et al. 2014; Carone et al. 2015; Carone et al. 2016;
Way et al. 2016). 
 Slow rotation weakens the Coriolis effect, and causes the atmospheric circulation to shift from 
a ``rapidly rotating'' regime characterized by zonal uniformity (banded cloud formation) and symmetry about the equator 
(e.g. like Earth presently), to a 
``slowly rotating'' regime characterized by day-side to night-side heat (energy)
 transport and circular symmetry about the substellar point.  The transition 
between these circulation regimes depends on the radius of the planet and rotation rate through the Rossby radius of deformation
(The Rossby radius of deformation is the length scale at which the Coriolis effect becomes important for determining the motion 
of an air parcel).  For an Earth 
sized planet, the Rossby radius of deformation exceeds the planet size when the rotational period reaches $\sim 5$ Earth days 
(Yang et al. 2014).  The climates of these slow rotating worlds are strongly influenced by the circulation regime, as convection, clouds, and 
horizontal transport are fundamentally altered (Yang et al. 2013; Kopparapu et al. 2016). On slow rotating planets, thick clouds at the 
substellar point tend to cool the planet by increasing the albedo. Thus, any attempt to accurately simulate the climate of planets around low 
mass stars, must take into account these inherently 3D effects, brought about by the slow and synchronous rotation of the planet.  These 
effects cannot be meaningfully parameterized in 1-D models used for studying habitable extrasolar planets 
(e.g. Kopparapu et al. 2013; Meadows et al. 2016).

Furthermore,  recent three-dimensional climate modeling studies predict that rapidly rotating Earth-like planets undergo a sharp 
transition between temperate and moist greenhouse climate states (Wolf \& Toon, 2015; Popp et al. 2016).  
 However, another study with a different model (Leconte et al. 2013a) found that cold stratosphere appears
to preclude an increase of stratospheric humidity, preventing a moist greenhouse state. 
Wolf \& Toon (2015) used a modified version of the NCAR CAM4 3-D climate model and found
stable, moist greenhouse solutions are indeed possible for an Earth-Sun configuration. Furthermore, Wolf \& Toon (2015)
 found comparable stratospheric temperatures (always $\ge 150$
K) to Leconte et al. (2013a) at low surface temperatures, but at high surface temperatures they found
much warmer (up to $\sim 210$ K) stratospheric temperatures and correspondingly higher
stratospheric $\h2o$ mixing ratios. In a third study, Popp et al. (2016) used the ECHAM6 3-D
climate model and found stable moist greenhouse solutions very similar in description to those
found by Wolf \& Toon (2015), albeit at lower stellar fluxes in their model. If the results of Popp
et al. (2016) and Wolf \& Toon (2015) are correct, then water would be lost from a planet's
surface well before a runaway greenhouse occurs, and possibly before surface temperatures can
rise above limits for habitability. Another study by Kasting et al.(2015), using a 1-D climate model,
 found that at higher surface temperatures the stratosphere warms and water-vapor indeed 
becomes a dominant component gas in the
atmosphere. Thus, contrary to Leconte et al.(2013a), catastrophic
water loss may be possible from a moist greenhouse planet.

Wolf and Toon et al. (2015) argue that the transition to a moist greenhouse is associated with a fundamental change to the radiative-convective 
state of the atmosphere.  For a rapidly rotating Earth-like planet around the Sun, if global mean temperatures 
exceed $\sim 310$ K then increasing water vapor causes both strong solar absorption and inefficient radiative cooling 
in the low atmosphere,  resulting in net radiative heating of the layers above the surface.  This heating forms a 
thermal inversion, which is stable against deep convection and thus prohibits the formation of convective clouds 
(See section 3.3 for more discussion). 

 Kopparapu et al. (2016) found that the above-described radiative-convective 
transition also occurs on slow and synchronously rotating Earth-like planets, which are expected around low mass stars.  
While rapidly rotating planets can maintain climatological stability beyond this transition due to cloud adjustments in the 
upper atmosphere, this transition is catastrophic for planets located near the inner edge of the HZ around low 
mass stars.  Slow-rotating planets are effectively shielded from the host star by thick convectively produced 
clouds located around the substellar point.  These planets can remain habitable despite incident stellar fluxes up to twice 
that of the present day Earth (Yang et al. 2014;  Way et al. 2016; Kopparapu et al. 2016).  However, the radiative-convective transition and 
subsequent onset of the near surface inversion stabilizes the substellar atmosphere, and thus the convective cloud deck 
rapidly dissipates.  Even a small dent in this substellar cloud shield then lets in a tremendous amount of solar radiation, 
destabilizing climate towards an immediate thermal runaway.  

Here, our goal is to study terrestrial planet atmospheres near the inner edge of the HZ around low mass stars 
($2600$K $\le T_{eff} \le 4500$K), using a state-of-the-art 3-D climate system model, with updated and validated 
$\h2o$ radiative transfer. We provide new estimates on the inner edge of the HZ for low mass stars, while differentiating 
between moist greenhouse and runaway greenhouse limits to planetary habitability. Our results indicate that, even for 
synchronously rotating planets, the inner edge of the HZ around low mass stars is significantly farther out 
(i.e. at lower stellar fluxes) than is predicted by earlier studies (Yang et al. 2013, Kopparapu et al. 2016).

Even more interesting is that several simulated planets in the slow rotation regime have water-rich atmospheres 
(mixing ratios $\sim > 10^{-3}$) that are within the classic ``moist-greenhouse'' regime, but maintain mild surface temperatures ($\sim 280$ K)
due to the thick cloud shield near the substellar point. In comparison, a 
moist-greenhouse in a 1-D climate model occurs at a surface temperature of $\sim 340$ K, 
and 3-D climate models show moist greenhouse atmospheres occurring with surface temperatures of $\sim 350$ K 
(Wolf \& Toon, 2015; Popp et al. 2016), in each case too hot for life similar to our’s  \citep{SH2010}.
 Here, we calculate that these water-rich habitable atmospheres are stable for 
hundreds of million or even several billion years. Depending upon the amount of water lost, it is
possible that some planets can smoothly transition to water-poor or dry planets, and can still retain habitability
in these dry states
\citep{Abe2011, Leconte2013b, Kodama2015}. We speculate on the number of such planets that can be detected by the {\it TESS} mission and
perhaps eventually be characterized by {\it JWST}, as these planets are much closer to the star than planets within the 
conventional HZ. 

The outline of the paper is as follows: In \S\ref{sec2} we describe
	the updates to our 3-D GCM  climate model.
	In \S\ref{results} we present results. In \S\ref{discussion}, we discuss the ramifications of our results on habitable zones 
and observations and conclude in \S\ref{conclusions}.

	\section{Model}
	\label{sec2}

\subsection{Atmospheric Model}
Here, we use a modified version of the Community Atmosphere Model version 4 (CAM4) from the National Center for Atmospheric Research in Boulder, CO (Neale et al. 2010).  We use $4^{\circ} \times 5^{\circ}$ horizontal resolution, with 40 verticals levels extending to a model top of 
$\sim 1$ mbar.  For all simulations we assume a planet with Earth radius, mass, and gravity.  The atmosphere is taken to consist only of 
$\n2$ and $\h2o$.  Other species such as O$_{2}$, O$_{3}$, CO$_{2}$, CH$_{4}$, and trace gases have been removed 
for simplicity, and habitable planets with active carbonate-silicate cycles should have low CO2 levels if their surfaces temperatures are hot.
  Thus, here our calculation of the inner edge of 
the habitable zone represents an innermost boundary.  The addition of other greenhouse gas species would serve only to push the inner edge 
of the HZ farther away from the star.   The total pressure of the atmosphere is given as 
$P_{tot} = P_{\h2o} + P_{\n2}$, 
 where $P_{x}$ is the partial pressure of each gaseous component. $P_{\n2}$  is set to 1 bar in all simulations, while atmospheric water 
vapor varies self consistently depending on the temperature and relative humidity of the atmosphere. 

We assume an aquaplanet (i.e. an entirely ocean covered planet with no land), with a 50-meter deep thermodynamic ``slab'' ocean
 (Bitz et al. 2012). Ocean heat transport is to set to zero everywhere.  While we focus on the inner edge of HZ,  many of our 
simulations still have ice cover.  We include sea ice using the Los Alamos sea ice model CICE version 4 (Hunke and Lipscomb, 2008).  We use the 
default albedo parameterization, which divides ocean, sea-ice, and snow albedos into two bands; visible and near-infrared divided at 
$0.7 \mu$m.  Following Shields et al. (2013) we set the visible (near-infrared) sea ice albedo to 0.67 (0.3), the snow albedo to 0.8 (0.68),
 and the ocean albedo to 0.07 (0.06) respectively. 

	We use the finite volume (FV) dynamical core (Lin and Rood, 1996).  The FV dynamical core has been modified to improve numerical 
stability (courtesy of C. Bardeen).  This entails incrementally applying physics tendencies for temperature and wind speed, evenly through out 
the dynamical sub steps.  Note, that several additional sensitivity tests were conducted using the spectral element dynamical core on a 
cubed-sphere grid (Lauritzen et al. 2014).  We use 
physics timesteps of 30 minutes with 48 dynamical substeps per physics timestep.  The radiative transfer is calculated once every 90 minutes.  
All simulations are initiated with surface temperatures matching that of the modern Earth.
Simulations are run until they enter a runaway state, or reach statistically steady state, which typically takes about 40 Earth years. 

\subsection{Radiative Transfer}

Previous calculations of the inner edge of the HZ for tidally locked planets around late K- and M-dwarf stars were conducted with
CAM3 and CAM4 (Yang et al. 2014; Wang et al. 2016; Kopparapu et al. 2016), using the native (i.e. unmodified) radiative transfer code.  
According to the scientific description of CAM\footnote{http://www.cesm.ucar.edu/models/atm-cam/docs/description/description.pdf},
 the long-wave radiative transfer is based on the broad-band model approach described by Kiehl and Briegleb (1991) and 
Kiehl and Ramanathan (1983).  The shortwave radiative transfer uses the $\delta$-Eddington approximation described in 
Briegleb (1992), across a coarse pseudo-spectral grid. Absorption by water-vapor is based on HITRAN2000 line database 
(Rothman et al., 2003), and incorporates the Clough, Kneizys, and Davies (CKD) 2.4 prescription for the continuum 
(Clough et al., 1989). 
Recently it was shown that the native CAM radiative transfer code significantly underestimates absorption by water vapor in both the longwave 
and in the shortwave (Yang et al., 2016).  This means that the inner edge of the habitable zone defined by Yang et al. (2014) and 
Kopparapu et al. (2016) are both placed too close to the host star.  

Yang et al. (2016) also indicate that the radiative transfer code first described by Wolf and Toon (2013) and later used by 
Wolf and Toon (2014, 2015) to study moist greenhouse atmospheres for Earth around the Sun, does well in the longwave against its reference 
model (LBLRTM), but absorbs too much solar radiation for warm ($T_{s} > 300$ K), saturated atmospheres.  Another result to emerge from 
Yang et al. (2016) is that spectral resolution in the near-infrared is critical for accurately calculating shortwave radiative transfer in 
water-rich atmospheres.  Here, we improve upon the radiative transfer scheme first described in Wolf and Toon (2013), by implementing a new 
correlated-k approximation using the latest spectral database, HITRAN 2012, and by increasing the spectral resolution in the near-infrared. 
New k-coefficients were created using HELIOS-K, a new open source spectral sorting program designed for graphics processing units 
(Grimm and Heng, 2015).  The utilization of massively parallel graphics processing units enables a several orders of magnitude increase in 
computational efficiency.  This allowed us to test numerous different spectral resolutions, and achieve an optimal balance of speed and 
accuracy for our project.  We settled on a 42 spectral interval grid ranging from 10 - 50,0000 cm$^{-1}$, with 8 Gauss points per interval.  
The shortwave stream is computed over 35 spectral intervals from 820 to 50,000 cm$^{-1}$.  The longwave stream is computed over 19 spectral 
intervals from 10 to 4000 cm$^{-1}$.  

We only consider absorption by $\h2o$, with the correlated k coefficients calculated for 
a pressure grid from 10 bar to 0.01 mb with successive pressure levels following 
log10(P1/P2) = 0.1, and a temperature grid from 100 to 500 K with $\Delta T = 25$ K. Furthermore, we have implemented a new water vapor 
continuum parameterization, following the formalism of Paynter and Ramaswamy (2011).  This continuum treatment is derived from laboratory 
measurements taken at temperatures appropriate for atmospheres expected near the inner edge of the habitable zone, and is stronger 
than the commonly used MT-CKD continuum. Finally, we have implemented direct-integration by mid-layer Planck functions to correct some 
unphysical sawtoothing of heating rates found near the model top in previous versions. We use the two-stream radiative transfer solver from 
Toon et al. (1989).  

 Cloud overlap is treated using the Monte Carlo Independent Column Approximation (MCICA), assuming maximimum-random overlap (Pincus et al. 2003).  MCICA uses stochastically generated sub-columns to represent sub-grid scale cloud variability.  Sub-columns are then randomly sampled across the spectral integration.  MCICA is specifically designed for use in 3D models, and is well tested in both GCMs (Barker et al. 2008) and numerical weather prediction models (Hill et al. 2015).

To test our code we follow the methods of Yang et al. (2016), using clear-sky saturated atmospheric profiles with surface temperatures ranging 
from 250 K to 360 K, and 200 K isothermal stratospheres.  For shortwave calculations we assume a 3400 K blackbody stellar spectrum as input. 
 In Fig. 1 we compare results from our correlated-k radatiave transfer code with that from a line-by-line code, LBLRTM (Clough et al. 2005) 
for both the longwave and shortwave.  Calculations of both outgoing longwave radiation (OLR) and absorbed shortwave radiation (ASR) remain 
within $\sim 3$ W m$^{-2}$ compared with LBLRTM for all atmospheric profiles tested.  This is accurate to within about $\sim 1 \%$ with respect 
to OLR and ASR.  Accuracy is a substantially improved compared to previous iterations of our correlated-k code and also compared to the 
native CAM radiative transfer (Yang et al. 2016). 

\begin{figure}[!hbp]
\centering
\includegraphics[width=.85\textwidth]{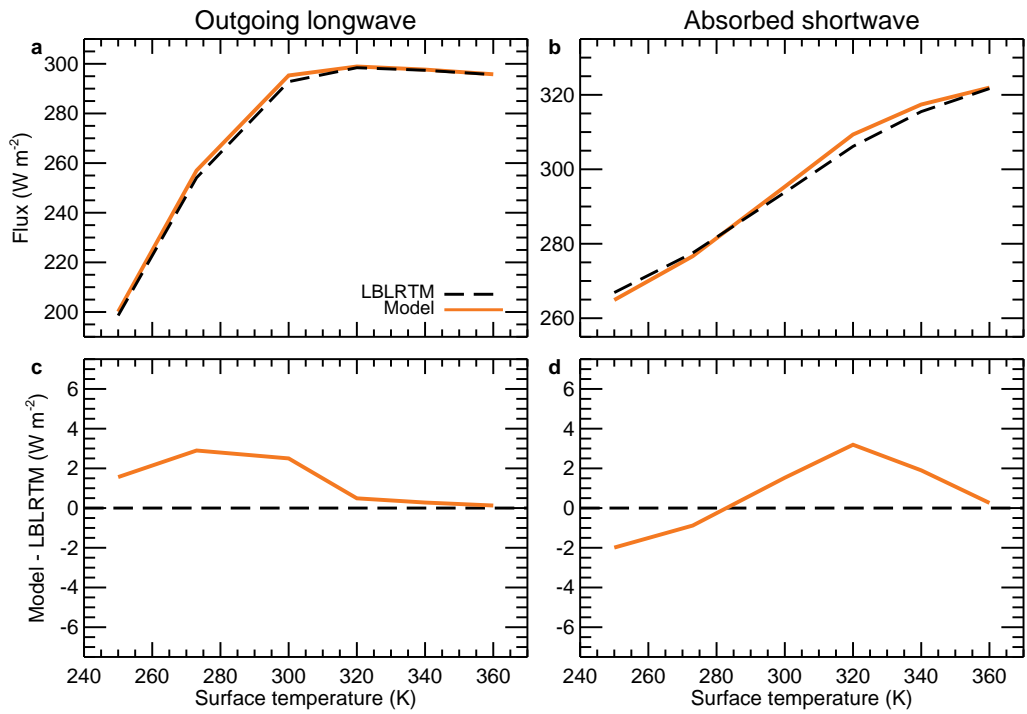}
\caption{Comparison of radiative transfer results from our GCM (solid curve) with that from LBLRTM (dashed curve). In total, there is 
$\sim 1\%$ (3 Wm$^{-2}$) difference between the two codes in both longwave (left) and shortwave (right).  }
\label{fig1}
\end{figure}

\subsection{Moist Physics}

Condensation, precipitation, and evaporation use the bulk microphysical parameterizations of Rasch and Kristjánsson (1998).  Deep convection 
is treated using the parameterization of Zhang and McFarlane (1995).  This scheme has been improved to include convective momentum 
transport (Richter and Rasch, 2008) and dilute entraining plumes (Raymond and Blyth, 1986, 1992). The deep convection scheme uses a plume 
ensemble approach, where convective updrafts and downdrafts occur wherever the low atmosphere becomes conditionally unstable.  The entropy 
closure calculation within the deep convection scheme has been updated (courtesy of C.A. Shields) to use the more robust numerical approach of 
Brent (1973).  A secondary convective parameterization treats shallow convective overturning at all layers in the atmosphere (Hack,1994).  
CAM4 self-consistently accounts for changing atmospheric pressures due to changing water vapor amounts in each grid cell. The total grid cell 
mass is determined from tendencies imparted by advection, convection, turbulent mixing, and large-scale stable condensation and evaporation. 
CAM4 uses virtual temperature corrections to account for variations in density and a similar approach accounts for the variation in specific 
heat between moist and dry air.

Cloud fractions are determined for three separate types of clouds. Marine stratus cloud fractions depend on the stratification of the 
atmosphere between the surface and the 700 mbar level. Convective cloud fractions depend on the stability of the atmosphere and the 
convective mass flux. Layered clouds at all altitudes depend on the relative humidity and the pressure level.  Cloud liquid droplet radii 
are assumed to be $14\mu$m everywhere in the model.  Ice cloud particle effective radii follow a temperature dependent parameterization and can 
vary in size from a few tenths to a few hundred microns.  A typical ice cloud in the model has an effective radius of $\sim 50\mu$ m at air 
temperatures of 240 K and $\sim 20\mu$m at air temperatures of 200 K.

\subsection{Stellar properties}

We conduct simulations for planets around stars with effective stellar temperatures of $2600$ K, $3000$ K, $3300$ K, $3700$ K, $4000$ K, and 
$4500$ K.  We assume that all of our planets are in synchronous rotation with their host stars.  Leconte et al.(2015) argue 
that planets are expected to have a nonsynchronous rotation if they are in the habitable zone of stars more massive
than $\sim 0.5$ to $0.7M_{\odot}$ (depending on their location
in the habitable zone). Our stellar properties remain below this limit.  The orbital period and thus the planetary 
rotation rate of each case is calculated self-consistently with Kepler's laws, following equation 3 in Kopparapu et al. (2016).  The stellar 
effective temperature, mass, luminosity, stellar flux at the planet, and the planetary rotation rate are strictly intertwined.
 We repeat the equation from Kopparapu et al.(2016), as this is critical in choosing our rotational periods of planets in
our simulations:

\begin{eqnarray}
P (years) &=& \left[\left(\frac{L_{\star}/L_{\odot}}{F_{P}/F_{\oplus}}\right)^{3/4}\right] \left[{M_{\star}/M_{\odot}}  \right]^{-1/2}
\label{finalp}
\end{eqnarray}

 where $P$ is the orbital period of the planet (for synchronous rotation, it is also the rotational period), $L_{\star}/L_{\odot}$ is the luminosity of the star with respect to the bolometric luminosity of the Sun, 
$F_{P}/F_{\oplus}$ is the stellar flux incident on a planet ($F_{P}$), compared to the flux on Earth ($F_{\oplus}$) and
$M_{\star}/M_{\odot}$ is the stellar mass in Solar mass units.
This equation indicates that the orbital period of a planet in synchronous rotation can be calculated from the luminosity, mass and the incident flux,
and it cannot be chosen irrespective of the stellar type. We then use a parameterized relationship between $T_{eff}$, $M_{\star}$, 
and $L_{\star}$ (as shown in Table 1 of Kopparapu et al. (2016)) to choose a specific stellar temperature. A list of
simulations with stellar and planetary characteristics shown in Figures 3, 4 \& 5 is tabulated in Table 1.

	For stars with $T_{eff} > 3200$ K, we use the empirically derived relationships between the mass, luminosity, and effective 
temperature of the host stars from Boyajian et al. (2013).  These relationships are valid for stellar metallicities ranging from 
[Fe/H] = -0.5 to +0.1 dex.  Note that Boyajian et al. (2013) conclude that metallicity only affects the color index of a star, 
however metallicity does not appear to impact the global properties of star: temperature, radius, and luminosity.  
Stars with $T_{eff}  < 3200$ K fall outside the validity range for the fits provided by Boyajian et al. (2013).  Thus, in this regime we use 
the stellar properties from the Baraffe et al. (2002) stellar evolutionary models, assuming an age of $\sim 3$ Gyr (log t(years) = 9.50). 
 For all cases, we use stellar spectra from the BT-SETTL grid of models (Allard et al. 2003, 2007), assuming a stellar metallicity of 
[Fe/H] = 0.0.  We conduct simulations for planets under increasing stellar fluxes, until a runaway greenhouse is triggered.

\subsection{The SMART Radiative Transfer Model}
The Spectral Mapping Atmospheric Radiative Transfer Model (SMART) is a versatile line-by-line fully multiple scattering 1-D radiative transfer 
model (Meadows \& Crisp 1996), capable of simulating a wide variety of planets. It has been validated against solar system objects (Robinson et al. 2014; Arney et al. 2014), and has been used to study a wide variety of types of atmospheres (Arney et al. 2016, Schwieterman et al. 2016, Charnay et al. 2015, Robinson et al. 2011). SMART has been modified to calculate transit spectra, including the effects of refraction (Misra et al. 2014).  Transit spectra can be generated from GCM results by using atmospheric columns from the planetary terminator region. 

\subsection{PandExo: A {\it JWST} simulator}
We use PandExo, a publicly availalbe simulator for {\it JWST} and {\it HST}. PandExo is described extensively in Batalha et al. 2017, but we 
briefly describe it here. PandExo relies on STScI's exposure time calculator Pandeia (Pontoppidan et al. 2016) to compute pixel level 
simulations based on the most up-to-date instrument characterization data. Therefore, it includes simulations of throughputs, background noise, 
read noise, point-spread-functions, optical paths and saturation levels. However, it does not fully simulate detector systematics, which have 
been suggested to create a noise floor of 20-30 ppm for the near-IR instruments and 50 ppm for MIRI (Greene et al. 2016). It also does not 
simulate any time varying noise sources such as stellar variability or spacecraft jitter and drift, which could add to {\it JWST}'s systematic 
noise floor (Barstow et al. 2015).

	\section{Results}
	\label{results}

\subsection{Control Simulations}

First we compare several sets of idealized simulations of synchronously rotating planets in order to determine the effect on climate of our updates to the 
radiative transfer code.   In Fig. 2 for all simulations we assume a 3400 K blackbody stellar spectra,
 and we fix the planetary rotation to a 
60-day synchronous orbit for all values of the incident stellar flux.  
 The blackbody spectra and fixed 60-day rotation rate were chosen to specifically to facilitate comparison with Yang et al. (2013).
We also assume a completely ocean covered world with no ocean heat 
transport.  Here, we compare CAM simulations using our new radiative transfer code, along with several results that use the native CAM 
radiative transfer scheme found in versions 3 and 4 of the model (Briegleb 1992).  Yang et al. (2013) used CAM3 with the spectral 
Eulerian dynamical core, Kopparapu et al. (2016) used CAM4 with the finite volume dynamical core, and we additionally test CAM4 with the 
native radiative transfer scheme and the spectral element dynamical core on a cubed sphere grid (Lauritzen et al. 2014).  Results from these 
three configurations, all using the native CAM radiative transfer (dashed lines), tend to cluster together irrespective of the choice in 
dynamical core and also independent of differences in the assumed planetary properties.  Note that Yang et al. (2013) simulations assume a 
planet of radius 2 R$_{\oplus}$ and a surface gravity of 1.4g, while the others assume 1 Earth mass and Earth-like gravity. 

 Note that some past studies have found that the CAM finite volume dynamical core failed to capture the upper atmosphere superrotating winds of Titan  and Venus (Parish et al. 2011; Lebonnois et al. 2012; Larson et al. 2014).  However, here we find no resultant differences in surface climate for slow rotating Earth-like planets using all three dynamical cores available in CAM.   We are currently preparing an accompanying manuscript that focuses on the atmospheric dynamics.

\begin{figure}[!hbp]
\centering
\includegraphics[width=.80\textwidth]{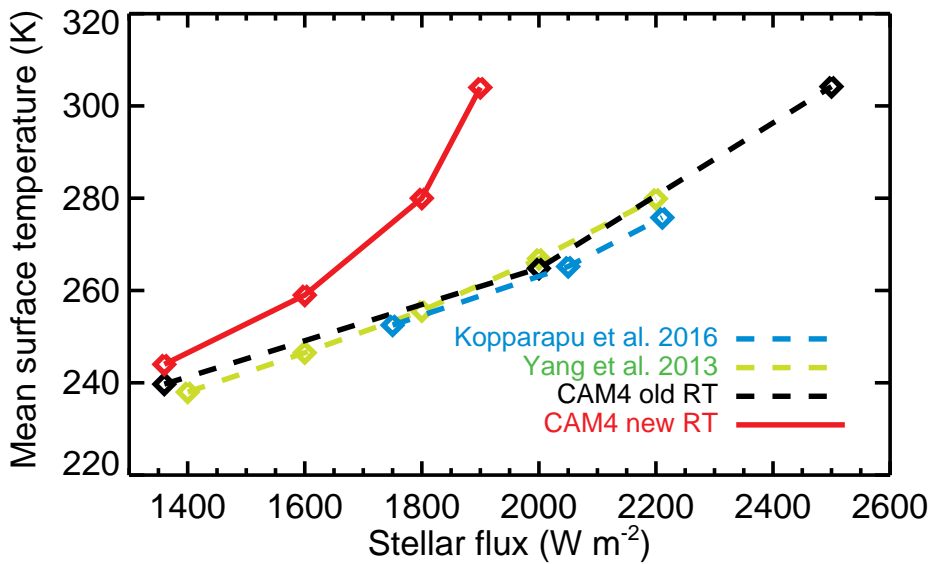}
\caption{Comparison of CAM simulations with the native band model radiative transfer scheme (dashed) and with our correlated-k radiative 
transfer scheme including updated $\h2o$ coefficients (red solid). Our new model indicates that a runaway greenhouse occurs  
at lower stellar fluxes compared with simulations that use
that used an older version of the radiative transfer. All the 
simulations assumed a $60$ day synchronously rotating planet in orbit around a $3400$K M-dwarf star.}
\label{fig2}
\end{figure}

Simulations with our updated radiative transfer (solid line) use the finite volume dynamical core.  These simulations yield similar 
temperatures under the present day Earth stellar flux.   However, as the stellar flux is increased, simulations using our updated radiative 
transfer code increase in temperature much more rapidly.  For these control simulations, a runaway greenhouse is clearly diagnosed for stellar fluxes 
exceeding $1900$ Wm$^{-2}$ using our updated code, while using the native radiative transfer a runaway greenhouse is not found until stellar 
fluxes exceed $2500$ Wm$^{-2}$.  In our new simulations, substellar clouds are still protective against high stellar fluxes, as first shown 
by Yang et al. (2013), however, the stronger longwave and shortwave absorption of our model causes the the inner edge of the HZ to be shifted 
to lower stellar fluxes.  This result comes as no surprise in light of the work of Yang et al. (2016), and the results shown here in Fig. 1.

\subsection{Habitable moist greenhouse states}

We next conduct a suite of simulations for ocean covered planets around late K- and M-dwarf stars with effective temperatures of 2600 K, 
3000 K, 3300 K, 3700 K, 4000 K, and 4500 K.
Note the 2600 K star is roughly the same temperature as TRAPPIST-1, while Proxima Centauri is about 3000 K.
  Our primary purpose for these numerical experiments is to determine the inner edge of the HZ for 
both runaway and moist greenhouse limitations.  For each simulation set, we study climates with mean surface temperatures ($T_{s}$) that 
range from $\sim 255$ K up to the triggering of a runaway greenhouse, characterized uncontrolled warming (see section 3.3 for more details). 
 The stellar flux and planetary rotation rate are self-consistently calculated for each cases following section 2.3.  Our results are 
summarized in Fig. 3 and Fig. 4.

\begin{figure}[!hbp]
\centering
\includegraphics[width=.80\textwidth]{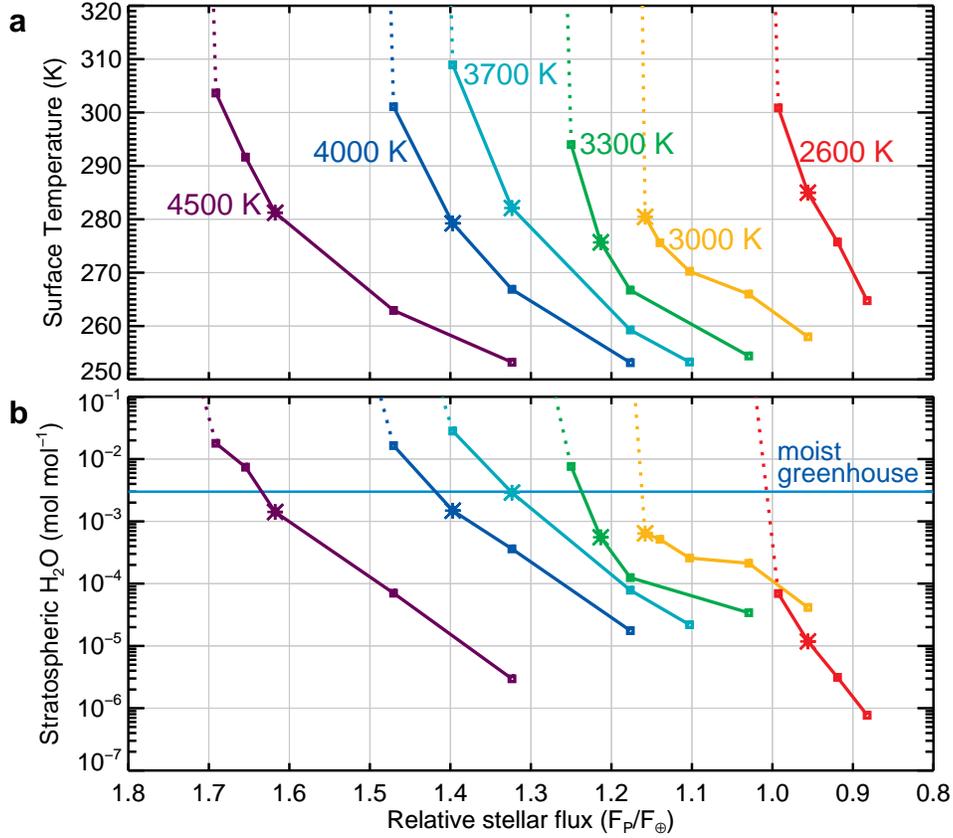}
\caption{Global mean surface temperature ($T_{s}$, top panel) and stratospheric $\h2o$ mixingratio 
(bottom panel) as a function of incident flux and stellar type. Here $F_{\oplus} = 1360$ W.m$^{-2}$.  Points connected by solid curves indicate stable climate simulations, 
and dashed lines
indicate the transition to a runaway greenhouse regime.  The ``$\star$'' sysmbol on the curves indicates those simulations shown in Fig. 8. The consistent shift of the curves towards lower stellar fluxes from 
K to M-dwarfs
is a result of shift in the stellar spectral distribution towards the red, and shorter orbital periods that will result in inefficient
substellar cloud albedo. The Horizontal blue line in panel b indicates the $\h2o$ mixingratio where 
the classical 
``moist-greenhouse'' limit occurs in 1-D climate models (Kasting et al. 1993). These two panels indicate that classical water-loss
limit occurs at significantly lower surface temperatures ($\sim 280$ K) compared to the predictions of Kasting et al.(1993) 1-D model 
($\sim 340$ K) for planets orbiting low-mass stars. }
\label{fig3}
\end{figure}

 Fig. 3a shows the global mean surface temperature for each simulation versus the relative stellar flux (i.e. the ratio of the incident 
stellar flux on the planet to that received by Earth presently, 1360 W m$^{-2}$).   All climatologically stable simulations are shown as 
points connected by solid lines labeled for each given host star.  Dashed-lines extending upwards indicate where a runaway greenhouse is 
triggered and climate warms uncontrollably despite only a small additional stellar forcing.  Similarly, Fig 4a shows $T_{s}$ as a function 
of the orbital period of the planets.  Recall that the orbital period and incident stellar flux are inextricably linked through the mass and 
luminosity of the host star (Kopparapu et al. 2016).  Furthermore, the orbital period equals the rotational period of the planet for 
synchronous rotators, as are 
simulated here.  Orbital periods for planets near the inner edge of the HZ are proportional the effective temperature of star, and range 
from $\sim 80-100$ days around a 4500 K star, to only $\sim 4$ days around a 2600 K star.
Our work suggests that TRAPPSIT-1 d, which receives about 1.1 times current Earth flux,  and  whose orbital period is about 4 days,
 is likely in a runaway greenhouse state.

 Our results also agree with Leconte et al. (2013b) who simulated synchronously rotating water-limited planets.   
If water is readily available on their surfaces, planet Gl 581c at 2.5 $S/S_{0}$ around a $\sim 3480$ K star, and 
planet 
HD85512 b at 1.85 $S/S_{0}$ around a 4715 K, have both likely undergone a runaway greenhouse.

 The consistent shift in the simulated planet curves towards the lower incident flux as we go to cooler stars
arises mainly due to 
(1) the shift in the stellar energy distribution towards the red part of the spectrum, where water-vapor
 is a strong absorber. Hence, a water-vapor rich planet around a cooler star does not need larger incident flux to quickly warm and
transition to moist or runaway greenhouse regime (2) the rotational periods of planets are shorter around M-dwarf stars compared
to K-dwarfs, resulting in less efficient substellar cloud albedo which warms the planet at lower incident fluxes.   

\begin{figure}[!hbp]
\centering
\includegraphics[width=.80\textwidth]{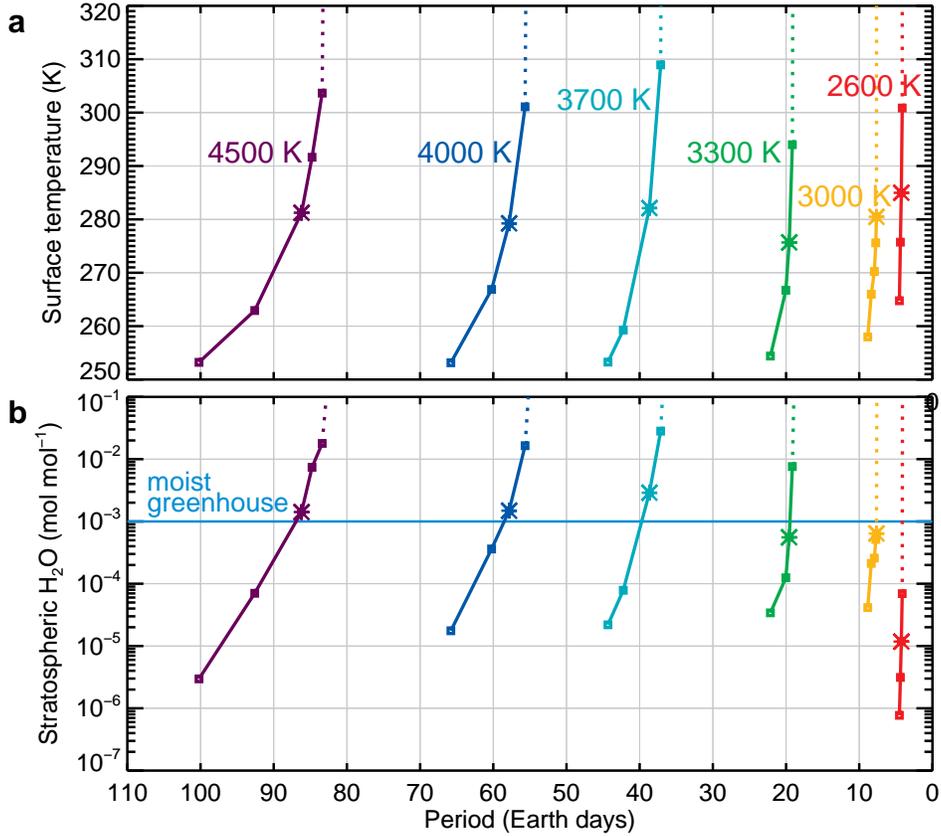}
\caption{Similar to Fig. 3, but $T_{s}$ and $\h2o$ mixingratio as a function of orbital period of the planet. For synchrnously rotating
planets, this is also the rotational period.}
\label{fig4}
\end{figure}

Fig. 3b and Fig. 4b show the stratospheric water vapor volume mixing ratio as a function of relative stellar flux and orbital period 
respectively.  Stratospheric $\h2o$ is taken here as the value at the top of the model (1 mb pressure).  The horizontal blue line in Fig. 3b
 and Fig. 4b indicates the classical ``moist greenhouse'' or water-loss limit to planetary habitability, used by Kasting et al. (1993) 
and Kopparapu et al. (2013).  This assumes that the total planetary $\h2o$ inventory is equal to the amount of water in Earth's oceans 
presently: $1.4 \times 10^{24}$ g of $\h2o$.  This equals $2 \times 10^{28}$ atoms cm$^{−2}$.   Water vapor that reaches the stratosphere is 
then photolyzed, releasing hydrogen, which escapes to space, thus irreversibly removing water from the planet.  If one assumes that escape is 
diffusion-limited (Hunten 1973), then the timescale for the loss of 1 Earth-ocean approaches the current age of Earth when the stratospheric 
water vapor volume mixing ratio is $\sim 3 \times 10^{-3}$ (Kasting et al. 1993; Kopparapu et al. 2013).   This is the classical criterion of a 
moist greenhouse climate.

Perhaps the most interesting result evident from Fig. 3 and Fig. 4 is that conditions necessary for both significant water loss and for the 
triggering of a runaway greenhouse occur at relatively low $T_{s}$.  The warmest stable climate against a runaway greenhouse is 
$\sim 310$ K, found for a planet around a 3700 K star, and is only $\sim 280$ K for planets around a 3000 K star.  Furthermore, moist 
greenhouse conditions become evident at mean surface temperatures as low as $\sim 280$ K.  Slow and synchronous rotators can have relatively 
large amounts of water in their atmospheres for a given $T_{s}$ due to strong and persistent deep convection on the substellar hemisphere that 
continuously lofts water vapor high into the atmosphere.  Thus terrestrial planets may lose their water to space while remaining relatively 
cool. This is in sharp contrast to simulations of  Earth around the Sun, where water loss conditions are not typically found until 
$T_{s} > 350$ K (Wolf and Toon, 2015; Popp et al. 2016).  Temperatures this warm, while able to support liquid surface water and extremophile 
life, could not support human biology (Sherwood \& Huber, 2010).   However, in this study, synchronously rotating planets around low mass stars 
reach moist greenhouse threshold  (water-vapor mixingratio $\sim 10^{-3}$ for 1 bar background N$_{2}$ atmosphere) while maintaining robustly habitable surface conditions.   A recent paper by Fuji et al.(2017) obtains a similar result using 
a different GCM (ROCKE-3D ,Way et al. 2017), broadly consistent with our results. Note, stable moist greenhouse climates are 
only possible for planets orbiting host stars with $T_{eff} \ge 3300$ K.   Planets around 3000 K and 2600 K stars, similar to Proxima Centauri and
TRAPPIST-1, bypass a water loss phase 
and proceed directly to a thermal runaway,  for a 1 bar background N$_{2}$ atmosphere atmospheree.

The low surface temperatures for the moist-greenhouse regime found in this study raises unexpected and interesting prospects for planets near 
the inner edge of the HZ around low-mass stars.  Low-mass stars may support habitable moist greenhouse worlds that could persist for several 
billion years during the water loss process before becoming desiccated.  Fig. 5 shows the timescale to lose 1 Earth ocean of water, 
calculated for each simulation based on diffusion-limited escape rates.   For 4500 K, 4000 K, 3700 K and 3300 K stars, ocean loss timescales 
can be $> 1$ Gyr for stable climates near the inner edge of the HZ.  This opens the intriguing possibility that such worlds could transition 
from water-worlds to dry planets while surface temperatures continuously remain within habitable ranges (Kodama et al. 2015).
  Ocean loss timescales from stable 
climates around 3000 K and 2600 K stars remain greater than 10 and 100 Gyr respectively.  Around any given star, once a runaway greenhouse 
instability is triggered, temperatures and thus atmospheric water vapor increase uncontrollably.  Then water would be lost from the planet in 
only a few million years, and the surface would be sterilized.

\begin{figure}[!hbp]
\centering
\includegraphics[width=.80\textwidth]{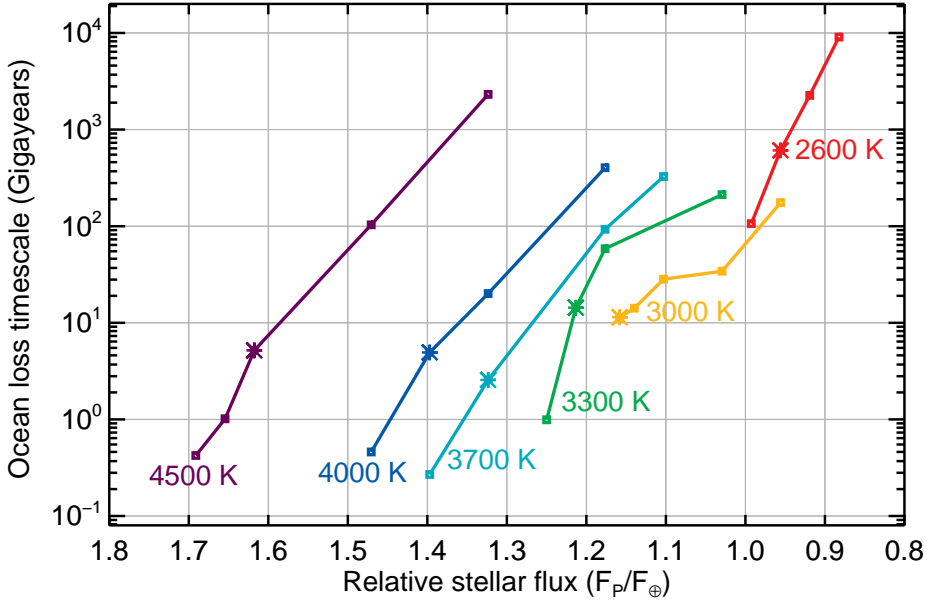}
\caption{Ocean-loss timescales as a function of incident stellar flux and stellar type. For planets that undergo moist-greenhouse state around
low-mass stars, the ocean-loss timescales can be $>1$ Gyr while maintaining lower surface temperatures ($\sim 280$ K). }
\label{fig5}
\end{figure}

\subsection{Transition to runaway greenhouse}

	In our simulations we can definitively identify the triggering of a runaway greenhouse.  Fig. 6 shows time series of temperature, 
planetary albedo, and the top-of-atmosphere (TOA) energy balance for several stable simulations, and for one simulation that enters a runaway 
greenhouse.  While numerical limitations of the model prevent us from exploring the end-state of the runway process where 
$T_{s} \sim 1600$ K (Goldblatt et al. 2013), still we can mark where the energy imbalance grows large and temperatures begin to rise 
uncontrollable.   This is sufficient for constraining the inner edge of the HZ for these worlds.  Fig. 6 shows 
time series model outputs only for simulations 
around a 3700 K star, however similar temporal evolution is found for each stellar type studied here.   In this case, climatologically 
stable simulations maintain a planetary albedo of $\sim 0.45$, and the TOA energy balance naturally approaches zero.  The runaway simulation 
experiences a collapse of the planetary albedo (dropping to $\sim 0.1$), coincident with a ballooning of the TOA energy balance.  At the time 
of model termination, the global mean surface temperature exceeds 350 K, and the maximum atmosphere temperature exceeds 400 K, and the TOA 
energy imbalance has risen to nearly $ + 150$ Wm$^{-2}$.

\begin{figure}[!hbp]
\centering
\includegraphics[width=.80\textwidth]{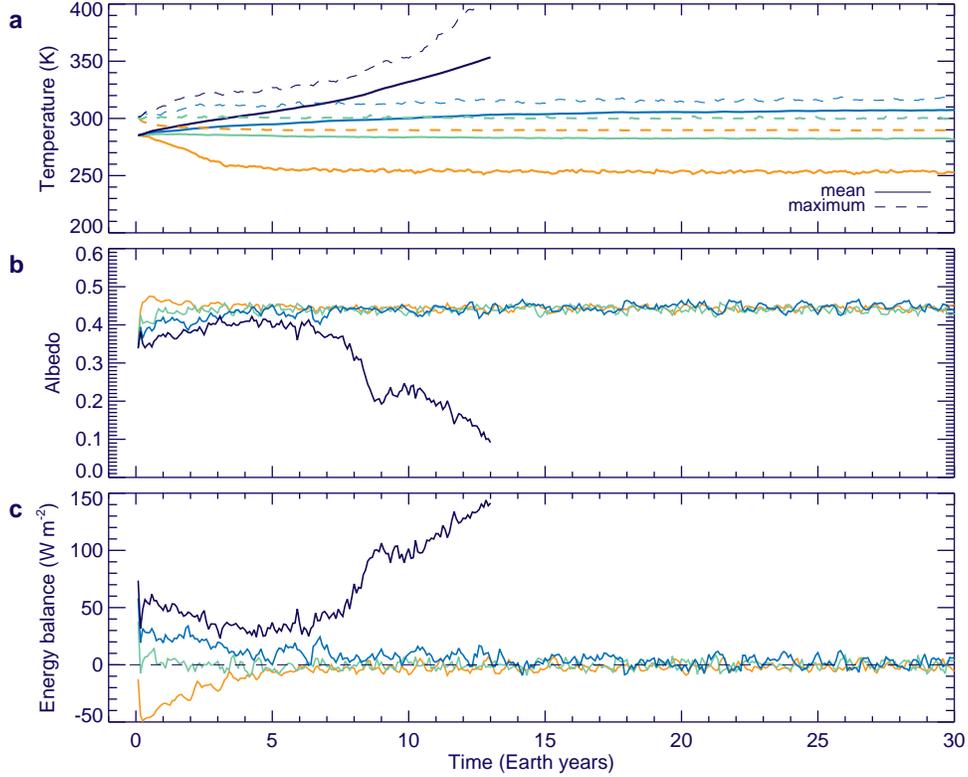}
\caption{Surface temperature (a), planetary albedo (b) and energy balance (c) for several planets around a 3700 K star. 
Note in panel (a) we show both the mean surface temperature and maximum atmosphere temperature. Both stable climate and
runaway simulations are shown. 
In stable simulations, the top-of-atmosphere energy balance goes to zero (c), as an equilibrium temperature (a) and 
albedo (b) is reached.  In the runaway simulation, the energy imbalance becomes large and temperatures begin to increases uncontrollably.}
\label{fig6}
\end{figure}

	For synchronously rotating planets around low-mass stars, the triggering of a runaway greenhouse is intimately linked to the thick 
substellar cloud deck.  Fig. 7 illustrates the time evolution of the vertical temperature structure and vertical cloud fraction of the 
atmosphere at the substellar point during the beginning stages of an ensuing runaway greenhouse.  The results in Fig. 7 correspond 
to the runaway greenhouse case also shown in Fig. 6.  

In the opening (Earth) years of simulation, the runaway case maintains thick clouds at 
the substellar point (Fig. 7b) while the planetary albedo remains large (Fig. 6b), similar to the stable cases.  However, as temperatures 
continue to warm, a radiative-convective transition occurs in the low atmosphere.  As described by Wolf \& Toon (2015), increasing water vapor 
in the atmosphere for warming climates causes the low atmosphere to become opaque to thermal radiation, while absorption of incident stellar 
radiation increases.  In total, this causes net radiative heating of the low atmosphere, which forms a strong temperature inversion (Fig 7a).
  The inversion encompasses the entire planet, including the substellar point, and stabilizes the low atmosphere against convection.  
 Inversion layers (such as the Earth’s stratosphere) are inherently stable against vertical mixing.
Without deep convection carrying moisture up from the boundary layer, the once thick and highly reflective convectively produced substellar 
cloud deck dissipates completely.  
 . High temperatures within the inversion layer, led to low relative humidities, and clear skies, despite a considerable water vapor 
burden
In the absence of these substellar clouds, the planetary albedo plummets and the planet is left 
unprotected against large stellar fluxes.  The energy imbalance skyrockets, and a thermal runaway becomes inevitable.  

\begin{figure}[!hbp]
\centering
\includegraphics[width=.80\textwidth]{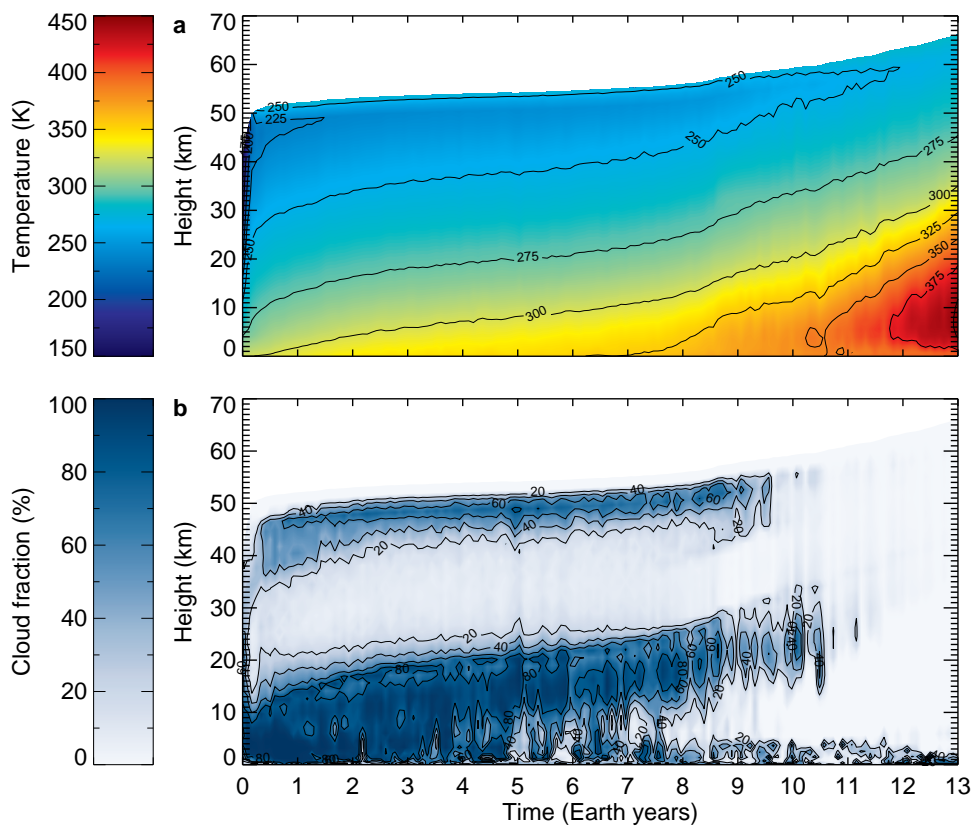}
\caption{ Diagnosis of a runaway greenhouse regime. The panels show the time evolution of vertical temperature (top panel) and vertical cloud
fraction (bottom panel) at the substellar point. Although during the beginning of the simulation there is a thick cloud cover at the 
sub-stellar point owing to deep convection carrying moisture up, the lower atmosphere stabilizes against convection as temperatures 
continue to warm. This shuts off convection and dissipates the cloud deck at the subs-stellar point, reducing the planetary albedo and 
energy imbalance ensues due to large incident stellar flux.}
\label{fig7}
\end{figure}

\subsection{Dependence on the host star}

The climates of planets in the HZs of low mass stars are inextricably controlled by the nature of the host star.  The orbital period, total 
incident stellar flux, and the stellar energy distribution are all correlated, determined by the mass and luminosity of the host star.   
Higher mass stars are bluer and brighter, and thus planets in their HZ must reside at further distances and thus will have longer periods and 
slower rotation rates.  Here we find that planets near the inner edge of the HZ around 4500 K stars will have periods of 
$\sim 80-100$ Earth days (Fig. 4).   Lower mass stars are redder and dimmer, thus HZ planets must remain closer to their host star and on 
shorter periods.  For instance,  near the inner-edge of the HZ planets around a 2600 K star have orbital periods of only $\sim 4-5$ days (Fig. 4).
  Differences in the 
rotation rate of these planets are of critical importance, because the rotation rate of the planet imposes a strong control on the atmospheric 
circulation regime through the modulation of the Coriolis force.  Slow rotation weakens the Coriolis effect, and causes the atmospheric 
circulation to shift from a ``rapidly rotating'' regime characterized by zonal uniformity and symmetry about the equator 
(e.g. like Earth presently), to a ``slow rotating'' regime characterized by day-side to night-side transport and circular symmetry about the 
substellar point.  The transition between these circulation regimes occurs when the rotation rate exceeds $\sim 5$ days for a $1$R$_{\oplus}$
  planet (Yang et al. 2014).  Thus planets near the inner edge of the HZ around a 2600 K star with a $\sim 4$ day rotation rate 
(like TRAPPIST-1d) should be in the fast 
rotating circulation regime.  Planets near the inner edge of the HZ around 3000 K star have rotations rates of $\sim 8$ days, and thus may 
represent a transitional state between rapid and slow rotators.  All other cases studied lie firmly within the slow rotating regime.  
The circulation regime critically feeds back upon the planet’s distribution of clouds.

\begin{figure}[!hbp]
\centering
\includegraphics[width=.85\textwidth]{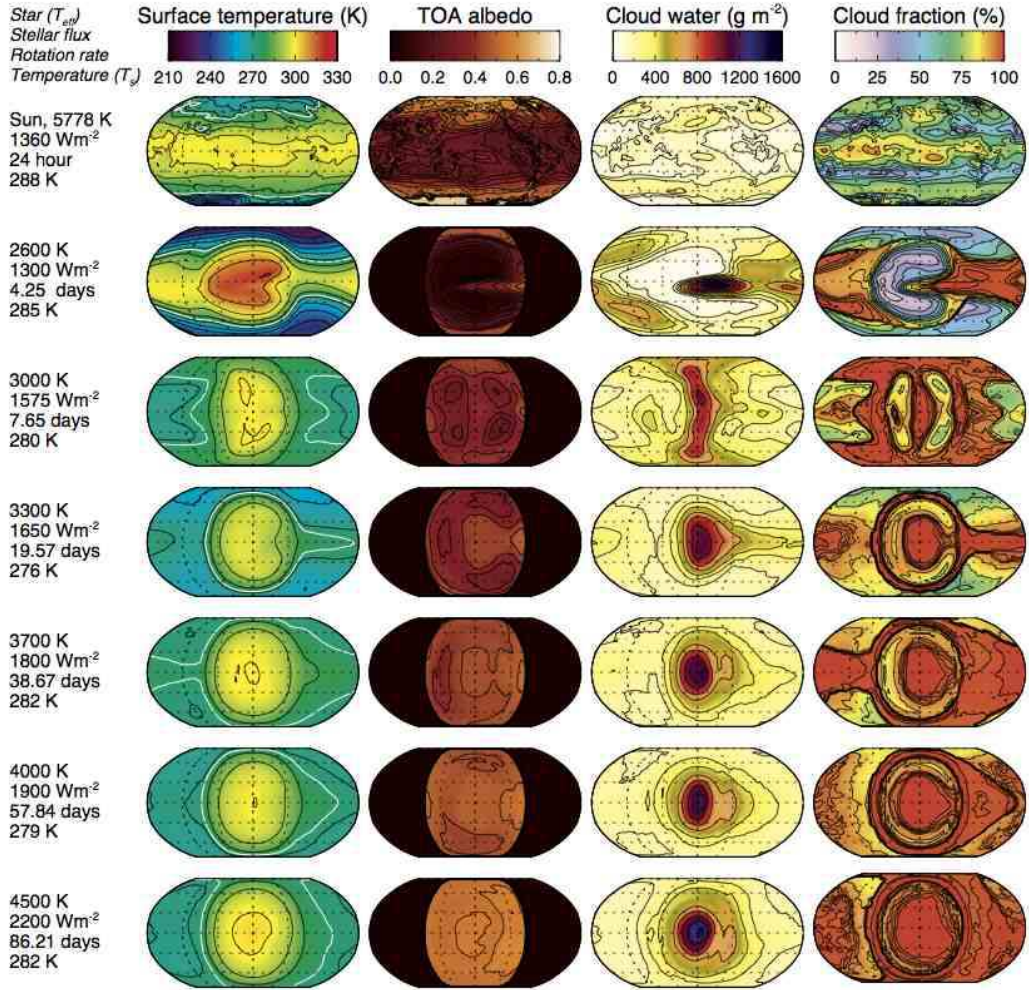}
\caption{From left to right in all rows: global mean surface temperature, TOA albedo, column integrated cloud water content and cloud fraction 
for planets that are represented by ``$\star$'' symbol in Figures 3, 4 \& 5. These results show that the climate of planets near the inner edge of
 the HZ around very low
mass stars ($\lesssim 2600$K) resemble the climate of fast-rotating planets like Earth.}
\label{fig8}
\end{figure}

Fig. 8 shows contour plots of surface temperature, TOA albedo, integrated cloud water column, and integrated cloud fraction for a present day 
Earth control simulation, and then for HZ planets around each of our 6 low mass stars studied here.  Note that the incident stellar flux on the 
planet, the planet's rotational period, and the planet's mean surface temperature are listed at the left margin of the Fig 8.  The white line 
over plot on the surface temperature contour indicates the location of the sea-ice margin.  The substellar point is located at the center of 
each panel.   All simulations shown in Fig 8 have moderate surface temperatures, with $276$ K $\le T_{s} \le 288$ K.  We have chosen to display 
simulations with similar $T_{s}$, because the mean temperature of the planet dictates sea-ice and the atmospheric water-vapor, both of which 
impose primary controls upon planetary climate.  By comparing simulations of nearly the same temperature (despite varying stellar flux, period, 
and spectra), we can normalize against temperature dependent feedbacks on climate.

	The climates shown in Fig. 8 are modulated by the rotation rate of each planet, through the feedback between the large-scale 
circulation and water clouds.  The TOA albedo is largely controlled by the fraction and thickness of the clouds. Note, in Fig 8 dark 
(light) colors represent areas of low (high) albedo for the planets in the second column.  The patterns emergent in planetary albedo mirror the distribution of the cloud water 
column and cloud fraction.  Regions of thick and ubiquitous clouds are naturally coincident with regions of high albedo.  Fast rotators, like 
 present day Earth and planets near the inner edge of the HZ around ultra low mass stars (i.e. 2600 K in this study), exhibit zonal patterns of temperature and clouds.
These climates feature the fewest clouds and the lowest albedos, and thus reach Earth-like temperatures at relatively low values of the incident
 stellar flux.  For planets around a 2600 K star, while the overall pattern is zonal, the planet retains the signature of a substellar hot spot
 due to its synchronous rotation.  Around increasingly higher $T_{eff}$ stars, the rotational period of HZ planets becomes necessarily longer and the
 circulation regime shifts.  Substellar clouds become increasingly thick and ubiquitous, resulting in higher planetary albedos, and larger 
required incident stellar fluxes to maintain Earth-like temperatures.  

\subsection{Atmospheric circulation}

The zonal circulation is the predominant circulation pattern on synchronously rotating planets, providing the primary means of energy 
transport from the substellar to antistellar hemisphere. Fig. 9 shows the mean zonal circulation (MZC, also known as the 
Walker circulation) for planets in synchronous rotation around stars with $T_{eff} = 3300$ K and $T_{eff} = 2600$ K (top row) as well as 
in units of pressure tendency).  Note that pressure tendency is a commonly used unit for expressing vertical motion in the atmosphere, i.e. a negative pressure tendency refers to rising motion, and vice-versa.

\begin{figure}[!hbp]
\centering
\includegraphics[width=.90\textwidth]{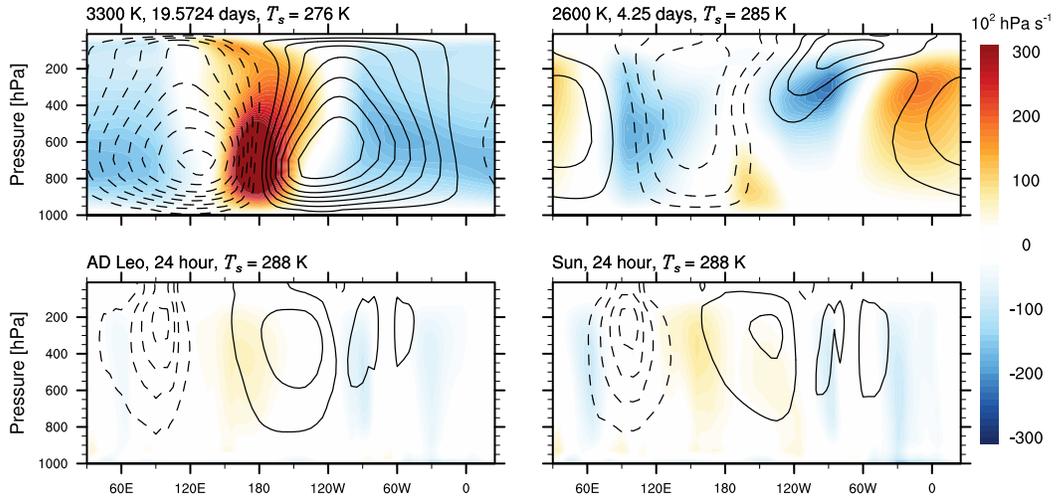}
\caption{ Mean zonal circulation (MZC) and vertical wind for synchronously rotating planets around 3300K and 2600K stars (top) and rapidly-rotating 24-hour period planets around AD Leo and the Sun (bottom). The contour interval for the MZC is 30×1011 kg s-1, with solid contours indicating clockwise circulation and dashed contours indicating counterclockwise circulation. Shading indicates pressure tendency, which corresponds to rising (warm colors) or sinking (cool colors) motion.}
\label{fig9}
\end{figure}

We note that a planet in a stable 4.25 day orbit near the inner edge of the HZ around a star with $T_{eff} = 2600$ K rotates fast enough that the Rossby deformation radius is 
less than the planetary radius. Thus, while the planet around the $T_{eff} = 3300$ K star is able to sustain strong rising motion at the 
substellar point and a hemisphere-spanning MZC from the day to night side, the planet around the $T_{eff} = 2600$ K star lacks a region of 
strong convection beneath its substellar point and shows a less organized structure to the time-average MZC.
Strong convection is also absent in the time-average MZC for the 24-hour rotating planets, where the lack of a fixed substellar point indicates
 the presence of a complex, diurnally-varying, MZC patterns closer to those observed on Earth today. Differences between the planet orbiting 
the Sun and Ad Leo are small, with planetary rotation rate being the prevailing factor for predicting differences in atmospheric circulation.

\subsection{Inner edge of the habitable zone}
Fig. 10a shows our simulated model cases around stars with $2600$ K $\le T_{eff} \le 4500$ K. The blue square
simulations have mild climates, with low stratospheric $\h2o$, whereas the green squares have
TOA water vapor of $10^{-3}$ or greater (``moist stratosphere''). Red crosses are simulations where the planet is in 
thermal runaway. For 
planets in synchronous orbits around stars with $T_{eff} > 3000$ K, there is an intermediate stage where the upper atmosphere
is dominated by water-vapor, 
before the planet transitions to a runaway with an increase in stellar flux. For stars with
$T_{eff} \le 3000$ K, the planet directly transitions to a runaway from a stable-mild climate. As discussed in previous sections,
this abrupt transition
to a thermal runaway, without undergoing a water-vapor dominated atmosphere, arises partly due to the shorter orbital periods (and 
hence faster rotation rates) which advects the persistent substellar cloud deck to the night-side, 
decreasing the planetary albedo and increasing the surface temperatures. This effect is compounded by the shift in the peak
wavelength of the stellar spectra towards the near-IR for low mass stars, where $\h2o$-vapor is a strong absorber. 
Our updated absorption coefficients for water-vapor further magnify this effect leading to the runaway greenhouse phase for
any further increase in the stellar flux.

\begin{figure}[!hbp]
\centering
\includegraphics[width=.90\textwidth]{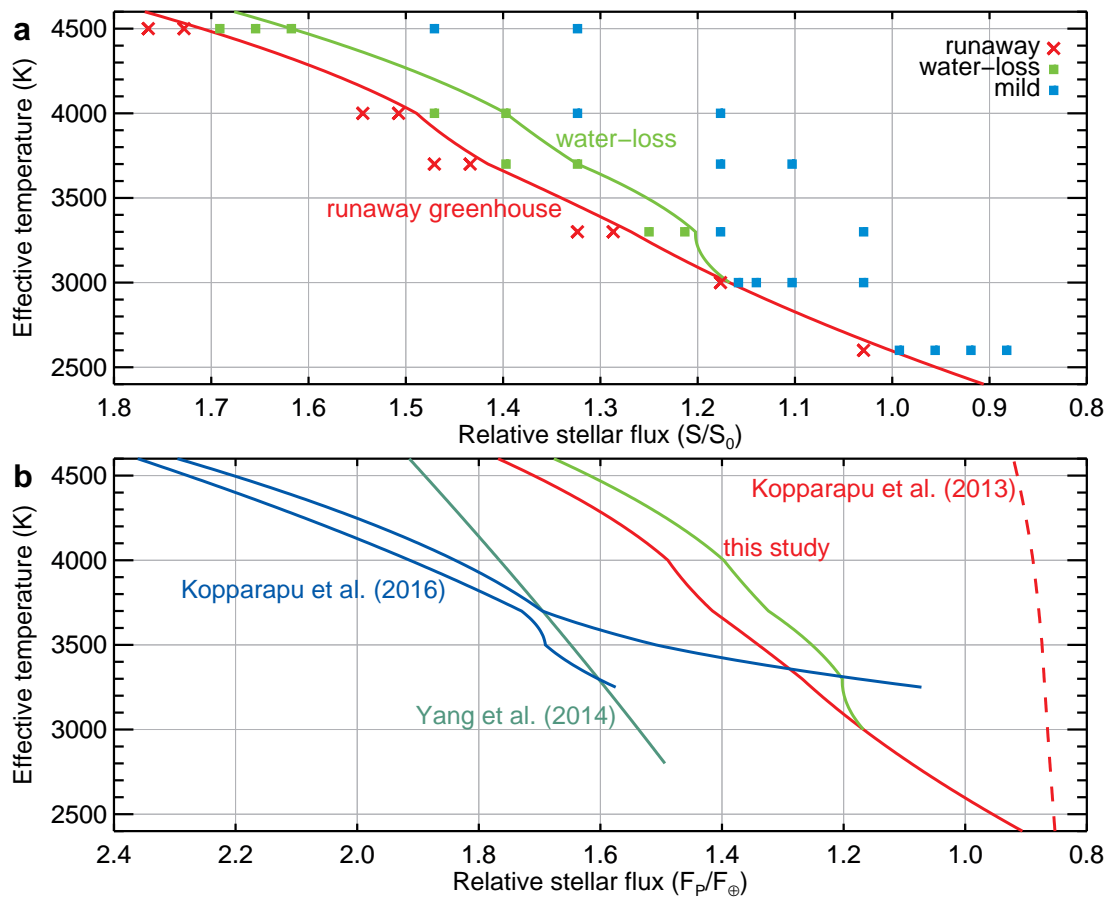}
\caption{Results of aquaplanet model simulations, after updating our GCM with new $\h2o$ absorption coefficients.
(a) For planets around stars with $\teff \ge 3000$K, there is an intermediate stage between mild climates (blue-squares) and
thermal runaway (red-cross), where the upper atmosphere
is dominated by water-vapor (green-squares) that could result in water-loss. Planets with $\teff <=3000$K directly transition
to runaway with an increase in stellar flux due to faster rotation rates, which advects the persistent substellar cloud deck to
the night-side,
decreasing the planetary albedo and increasing the surface temperatures.(b) Comparison of our new inner HZ results with previous
studies.}
\label{fig10}
\end{figure}

Fig. 10b shows our moist and runaway greenhouse limits in comparison with our previous study \citep{Kopp2016} and
\cite{Yang2014a}. Note that, as with \cite{Yang2014a}, \cite{Kopp2016} also assumed that the last converged solution 
represents the IHZ\footnote{For a detailed analysis of the last converged solution criterion, see section 3.4 of 
\cite{Kopp2016}.}. We assumed a solar metallicity for all the stars in this study. Compared to our previous results, the 
inner edge of the HZ (either moist or runaway GH) is located at considerably lower stellar fluxes for late-K and M-dwarf
stars. Considering that the global mean surface temperatures at the moist-greenhouse limit are $\sim 280$K, we conclude
that, for M-dwarf stars, the inner edge of the HZ is identified by the runaway greenhouse limit 
and moist greenhouse planets around these stars can be habitable, revising decades of previous assumptions about the 
moist greenhouse. 

 We remind the reader that our simulations assume a background N$_{2}$ atmosphere of 1bar. Several studies have shown that
the variation in the non-condensable gas fraction in the atmosphere (specifically the N$_{2}$ amount) can affect the OLR, 
and consequently the habitability limits through moist and runaway greenhouse 
\citep{Nakajima92, RayP2010, WP2014, Kopp2014}.  
For a higher N$_{2}$ amount in the atmosphere, the surface temperature needs to be higher to radiate the same amount of
OLR. The reason for this is that the amount of water-vapor fraction in the atmosphere is not high enough relative to the
N$_{2}$ fraction.  Consequently, planets with higher N$_{2}$ can potentially reach moist-greenhouse limits at higher
temperatures than $\sim 280$K. Conversely, if there is less than 1 bar $\n2$ in the atmosphere then water-loss could 
occur at even lower temperatures.  Still, our assumption of a 1 bar $\n2$ atmosphere is in line with previous studies 
(Kasting et al. 1993, Kopparapu et al. 2013a, Leconte et al. 2013a, Yang et al. 2014, Wolf \& Toon 2015, Popp et al. 2016).  An exploration of the effects of varying $\n2$ on moist greenhouse atmospheres in 3-D models is left for future 
parameter studies.

 We further remind the reader that differences exists across 3-D climates models of varying origin.  For example, 
outcomes for standard anthropgenic climate change scenarios can vary by several Kelvin over the next two centuries, 
depending upon the model being used (Rogelj et al. 2012).  However, the implied radiative forcings for simulated 
exoplanets are typically much larger, and yield wider spreads in model results (e.g. Popp et al. 2016, fig. 7).  
Formal model inter-comparisons are needed in order to make sense of these model differences.

\section{Discussion}
\label{discussion}
\subsection{Transit transmission spectral features}

\begin{figure}[!hbp]
\centering
\includegraphics[width=.55\textwidth]{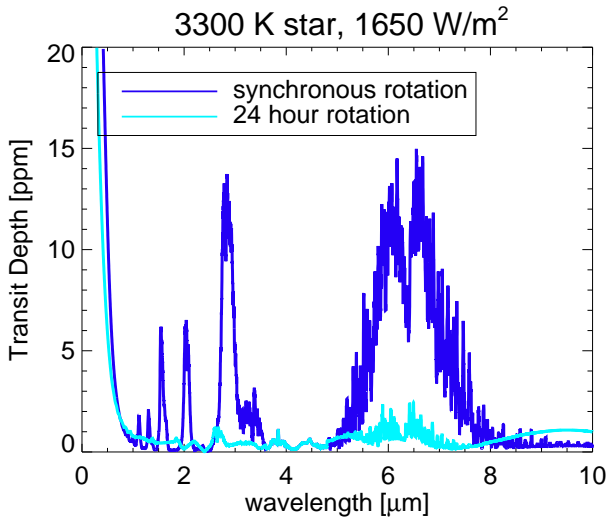}
\caption{Water vapor features are prominent in
the moist greenhouse atmosphere of an
Earth-sized planet orbiting a 3300 K star, with a global mean surface temperature of $\sim 285$K, 
when synchronous rotation is accounted for using GCMs (dark blue). Using an Earth-analog planet with 24-hour rotation
(light blue) around a M-dwarf will miss these water-vapor features in the spectra.}
\label{spec}
\end{figure}

M dwarf planets will be targeted by JWST, and if habitable planets can exist closer to these stars than the traditional HZ boundaries, their 
shorter orbital periods will allow us to observe them more frequently. For example, the TRAPPIST-1 system has three planets (d, e, and f) in
 the habitable zone. TRAPPIST-1 is visible for a total of 99 days in JWST's Cycle 1 (STScI APT).  Planet's d, e, and f have orbital periods of 
4, 6, and 9 days, respectively. Therefore in cycle 1 (1 year), Planet f can be observed for a maximum of 11 transit observations, 
whereas planet d can be 
observed for a maximum of 24 transits. 
TRAPPIST-1 has a stellar effective temperature of 2600 K so it is not directly applicable to the moist-greenhouse atmospheres discussed here.
 However, it does offer an estimate for what a feasible amount of observing time is for JWST. 
Therefore,  if we use TRAPPIST-1's observability window as a standard estimate, we can 
calculate the expected noise  on the transmission spectra of the planets of interest in this study
 (moist greenhouse planets around stars with $T_{eff}>3300$ K). 
In Fig. \ref{spec}, we show a simulated transit transmission spectrum from our GCM simulations, where the planet is undergoing moist-greenhouse around 
a M3 spectral type star with $T_{eff} = 3300$K, stellar radius of $0.137$ R$_{\oplus}$ (dark blue). All spectra were generated using the 
SMART model, a versatile line-by-line 1-D radiative transfer model (Meadows \& Crisp 1996).
 Transit spectra are generated from GCM results by using atmospheric columns from the planetary terminator region. 
This planet's atmosphere 
includes $\n2$ as the background gas with stratospheric $\h2o$ volume mixing ratio of $10^{-3}$ (see Figs. 3 \& 4 ), and self-consistent liquid and ice water 
clouds.
Fig. \ref{spec} shows that for a habitable moist greenhouse planet, there are two prominent water vapor features including one between 
5-8 $\mu$m, within the JWST MIRI LRS (Mid Infra-Red Instrument Low Resolution Spectrometer) bandpass, 
and another between 2.5-3 $\mu$m, within the NIRCam grism with the F322W2 
filter and the NIRSpec G235M/H bandpasses. These features, caused by a wet upper atmosphere, are a direct consequence of the strong and 
persistent convection that occurs at the substellar region for slowly rotating planets. 
 This strong substellar convection is effective at pumping water vapor into the high atmosphere.

In Fig. \ref{spec}, we also show a planet simulated around a M-dwarf, but with a standard 24-hour rotation period (light-blue spectrum).
 Previous studies of Earth-analog spectra for planets in the HZ around M-dwarfs have not included the impact of synchronous rotation 
(Barstow et al. 2015, Schwieterman et al. 2016). Taking into account the effect of M-dwarf rotation on cloud and species distributions 
will be crucial to accurately model the spectra of these worlds. Note how the planet simulated with a 24-hour rotation period results in a 
dramatically different spectrum in Fig. \ref{spec}  compared to including synchronous rotation. The water vapor spectral feature is 
$\sim 7$ times weaker in the atmosphere of the planet with the 24-hour rotation period because the strong substellar convection is absent.

	Despite the increased strength of the water absorption with the inclusion of synchronous rotation, 
the two strongest water spectral features in Fig. 11 are well 
below the postulated JWST noise floor of 20 ppm and 50 ppm for the nir-IR and mid-IR instruments aboard JWST (Greene et al. 2016). 
In reality, though, noise floors will not prevent observations of potentially habitable systems 
and the true noise floor of JWST remains to be determined. For example, TRAPPIST-1
is a comeplling enough target that it is already on JWST's Guaranteed Time Observer's plan. If other 
compelling small planets are found 
(planets that may fit the planet-type described here) they may end up being observed and so we should address what, 
if anything, we 
could learn from their transmission spectra. 

\begin{figure}[!hbp]
\centering
\includegraphics[width=.55\textwidth, angle=270]{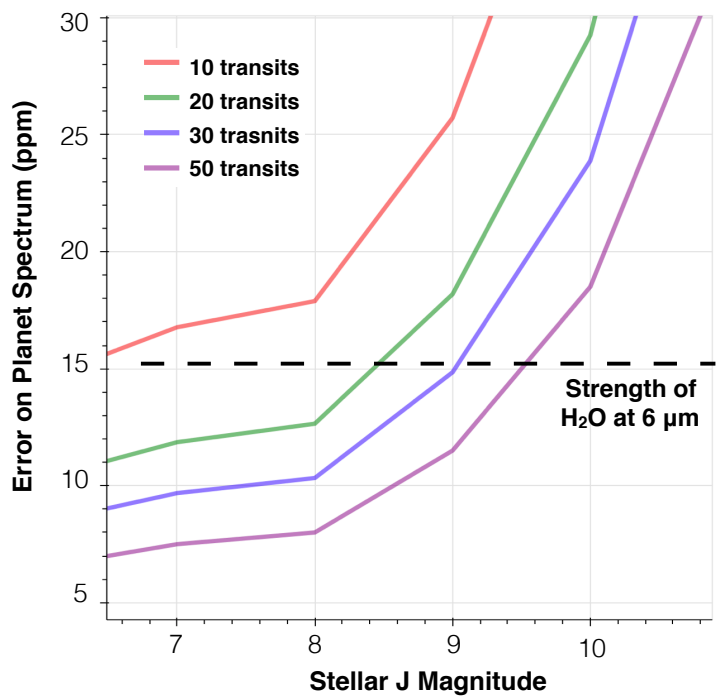}
\caption{Noise estimates calculated with PandExo at 6 $\mu$m   with JWST's MIRI instrument. The system simulated was a M3 stellar
spectral type with T= 3300 K. Each transit consists of a 2 hour in transit observation with a 2 hour out of transit baseline. At J=6, the 
duty cycle is calculated to be $33\%$ with 2 groups/integration. As the target gets dimmer, the duty cycle increases to $98\%$ at J=11 with 
130 groups/integration. No systematic noise floor has been added, although it is postulated to be $\sim 50$ ppm (Greene et al. 2016), 
based on comparisons with Spitzer and not measurements from the JWST MIRI team.
 The dashed line signifies the strength of the water absorption feature for the synchronously rotating system in Fig. \ref{spec}.}
\label{jwst}
\end{figure}

Assuming JWST encounters no systematic noise floor (which is unlikely), we can calculate how many observing hours would be 
needed to obtain sub-10 ppm precision. 
Noise floors are created when instrument systematics cannot be removed from coadding transits. Therefore, they create a minimum level equal to the strength of the systematic effect. 
Fig. \ref{jwst}  shows the approximate error on the spectrum of a MIRI observation at 6$\mu$m  for a 
range of stellar brightnesses (2MASS J), without any inclusion of a systematic noise floor. We chose the MIRI band because of the broadness of 
the $6 \mu$m water feature and because, as we are not including any systematic noise floor and observing an M3 star, the noise calculations are similar 
if the nir-IR simulations are binned to $R=100$. These calculations were performed using JWST's publicly available noise simulator, 
PandExo (Batalha et al. 2017) for the same planet system shown in Fig.\ref{spec}.  Assuming an observability window comparable to TRAPPIST-1 
and an inner edge planet, the planet system would need to be observed through the entirety of Cycle 1 (24 transits) to reach the approximate 
strength of the water feature. 

	Solely surpassing the strength of the water feature would yield a confirmation of the presence of water, not necessarily a constraint 
on the water abundance. In this particular scenario (aqua planet, no land), if the presence of water is confirmed, one could say the planet 
is either in a moist greenhouse or habitable state given it's stellar type and orbital period (see Fig. 4 ).
	If the system was then observed through Cycles $\sim 2-3$ ($\sim 60$ transits), a traditional Bayesian retrieval might be possible to 
constrain the water mixing ratio and corresponding temperature (such as that done in Barstow et al. 2016). Barstow et al. (2016) estimate 
retrieved errors on water abundance on the order of $\pm 1$ dex in mixing ratio. If it is possible to measure the water content to a 
level of precision of approximately $\pm 1$ dex, there would be three potential scenarios in accordance with Fig. 4 
(habitable, entering moist greenhouse, in moist greenhouse). Since transmission spectra will not probe the surface of terrestrial 
planets, we can use Fig. 4 to predict a surface temperature. This would require constraints on the stratospheric temperature, the water 
mixing ratio and information on the stellar type and orbital period.  

	In a scenario where by the end of Cycle 3, there is still no detection of any water spectral features, two possibilities exist: either 
the planet is a fast rotator so that strong substellar convection is absent, or the planet is dry. Both scenarios would be 
astrophysically interesting. 
Of course, there are other factors, which may complicate this simple qualitative view of exoplanet characterization. For example, here we have
 assumed an aqua planet with no land. The effect of adding continents, may also result in reduced $\h2o$ features. An exploration of these 
dependencies is left as future work.

\section{Conclusions}
\label{conclusions}
We have updated water-vapor absorption coefficients in our CAM 3-D GCM with HITRAN 2012 LBL database, and added continuum
absorption. Our results indicate, assuming synchronously rotating Earth-sized and Earth-massed planets with background 1 bar 
$\n2$ atmospheres, that the inner edge of the HZ for late-M to mid-Kdwarfs is further away from the star than 
previous studies. Furthermore, for stars with $T_{eff} \ge 3300$K, there appears to be a habitable moist-greenhouse regime,
which could not be predicted with 1-D climate models. For stellar $T_{eff} \le 3000$K, planets at the inner HZ directly
transition to runaway phase. We find that transition between synchronous and rapid-rotating planets occurs at an orbital
period of $\sim 4$ days, where the substellar cloud advected to the night side, decreasing the planetary albedo, and 
increasing the surface tempertures. 
We also find that using Earth-analog spectra for planets orbiting M dwarfs underestimates the strength of critical water-vapor feature because the 24-hour rotation period of 
Earth does not produce strong sub-stellar convection that transports water to the upper atmosphere. 
This strong convection is a feature of synchronously
 rotating planets as pointed out by several GCM studies, including this paper. Therefore, to interpret the spectral features of HZ planets
around M-dwarf stars, one needs to employ 3-D climate models to properly account for the effect of rotation on the atmosphere of
the planet.

	\acknowledgements

        The authors would like to thank  David Paynter for providing BPS coefficients. We also thank an
        anonymous reviewer for helpful comments.  
        R. K, E. T. W and J. H-M  gratefully acknowledge funding from NASA Habitable Worlds grant NNX16AB61G.
        R. K and G. A. also acknowledge funding from NASA Astrobiology Institute's  Virtual 
       Planetary Laboratory lead team, supported by NASA under cooperative agreement
       NNH05ZDA001C. N. E. B. acknowledges support from National Science Foundation under Grant No. DGE1255832.
       The Center for Exoplanets and Habitable Worlds is supported by the Pennsylvania State University, the Eberly 
       College of Science, and the Pennsylvania Space Grant Consortium. 
	This work was facilitated through the use of 
	advanced computational, storage, and networking infrastructure provided by the 
	Hyak supercomputer system, supported in part by the University of Washington eScience Institute.
        This work also utilized the Janus supercomputer, which is supported by the National Science Foundation 
        (award number CNS-0821794) and the University of Colorado at Boulder.
       Any opinions, findings, and conclusions or recommendations expressed in this material are those of the 
       author(s) and do not necessarily reflect the views of NASA or the National Science Foundation.

\begin{threeparttable}[h!]
\caption{Simulation cases shown in Figures 3-5. For a given stellar effective temperature ($T_\mathrm{eff}$),
the stellar luminosity ($L/L_{\odot}$), mass ($M/M_{\odot}$) and the incident stellar flux on the planet 
($F_{P}/F_{\oplus}$) are related with the orbital period $P$ through Eq.(1). We assume synchronously 
rotating planets, so the orbital period $=$ rotational period. The last column is the amount of
$\h2o$ available at the model top (1 mb), and also indicates the transition stages to water-loss and runaway
greenhouse limits.}
\vspace{0.1 in}
\centering
\begin{tabular}{|c|c|c|c|c|c|}
\hline
$T_\mathrm{eff}$ & $L/L_{\odot}$& $M/M_{\odot}$& $F_{P}/F_{\oplus}$& $P$(days)& Model top $\h2o$ mixingratio (1 mb)\\
\hline
$4500$ K & $0.189$ & $0.72$ & $1.323$ & $100.20$ & $2.97 \times 10^{-6}$\\
&&&$1.470$ & $92.59$ &$7.04 \times 10^{-5}$\\
&&&$1.617$ & $86.20$ & $1.41 \times 10^{-3}$ (water-loss)\\
&&&$1.654$ & $84.76$ & $7.41 \times 10^{-3}$\\
&&&$1.691$ & $83.37$ & $1.79 \times 10^{-2}$\\
&&&$1.727$ & $82.04$ & (runaway)\\
\hline
$4000$ K & $0.0878$ & $0.628$ & $1.176$ & $65.79$ & $1.76 \times 10^{-5}$\\
&&&$1.323$ & $60.23$ & $3.61 \times 10^{-4}$\\
&&&$1.397$ & $57.83$ & $1.48 \times 10^{-3}$\\
&&&$1.470$ & $55.65$ & $1.64 \times 10^{-2}$ (water-loss)\\
&&&$1.507$ & $54.63$ & (runaway)\\
\hline
$3700$ K & $0.0429$ & $0.520$ & $1.102$ & $44.33$ & $2.19 \times 10^{-5}$\\
&&&$1.176$ &$42.24$ & $7.85 \times 10^{-5}$ \\
&&&$1.323$ &$38.66$ & $2.89 \times 10^{-3}$ (water-loss)\\
&&&$1.397$ &$37.13$ & $2.82 \times 10^{-2}$ \\
&&&$1.433$ &$36.41$ &  (runaway)\\
\hline
$3300$ K & $0.00972$ & $0.249$ & $1.029$ & $22.12$ & $3.41 \times 10^{-5}$\\
&&&$1.176$ & $20.02$ & $1.24 \times 10^{-4}$\\
&&&$1.213$ & $19.57$ & $5.55 \times 10^{-4}$\\
&&&$1.250$ & $19.13$ & $7.60 \times 10^{-3}$ (water-loss)\\
&&&$1.280$ & $18.72$ & (runaway)\\
\hline
$3000$ K & $0.00183$ & $0.143$ & $0.955$ & $8.83$ & $4.14 \times 10^{-5}$\\
&&&$1.029$ & $8.35$  & $2.12 \times 10^{-4}$\\
&&&$1.103$ & $7.93$  & $2.57 \times 10^{-4}$\\
&&&$1.139$ & $7.74$  & $5.14 \times 10^{-4}$\\
&&&$1.158$ & $7.65$  & $6.39 \times 10^{-4}$\\
&&&$1.176$ & $7.56$  & (runaway, no water-loss limit)\\
\hline
$2600$ K & $0.000501$ & $0.0886$ & $0.882$ & $4.51$ & $7.71 \times 10^{-7}$\\
&&&$0.919$& $4.37$ & $3.13 \times 10^{-6}$\\
&&&$0.955$& $4.25$ & $1.18 \times 10^{-5}$\\
&&&$0.992$& $4.13$ & $6.94 \times 10^{-5}$\\
&&&$1.011$& $4.07$ & (runaway, no water-loss limit)\\
\hline
\end{tabular}
\label{table1}
\end{threeparttable}

\end{document}